\documentclass[runningheads]{llncs}
\usepackage{cmap}
\usepackage{xcolor}
\usepackage{graphicx}
\usepackage{listings}
\AtBeginDocument{\counterwithout{lstlisting}{section}}
\begin{document}
\title{AstroDS - A Distributed Storage for Astrophysics of Cosmic Rays. 
Current Status\thanks{Supported by the Russian Science Foundation, grant No.18-41-06003.}}
%
%
\author{
Alexander Kryukov\inst{1}\orcidID{0000-0002-1624-6131}
\and Igor Bychkov\inst{2}
\and Elena Korosteleva\inst{1}
\and Andrey Mikhailov\inst{2}\orcidID{0000-0003-4057-4511}
\and Minh-Duc Nguyen\inst{1}\orcidID{0000-0002-5003-3623}
} 
\authorrunning{A. Kryukov et al.}
%
\institute{M.V.Lomonosov Moscow State University, D.V.Skobeltsyn Institute of Nuclear Physics, Russia \\
\email{kryukov@theory.sinp.msu.ru} 
\and Matrosov Institute for System Dynamics and Control Theory, Siberian Branch of Russian Academy of Sciences, Russia 
}
\maketitle              
\begin{abstract}

Currently, the processing of scientific data in astroparticle physics is based on various distributed technologies, the most common of which are Grid and cloud computing. 
The most frequently discussed approaches are focused on large and even very large scientific experiments, such as Cherenkov Telescope Array. 
We, by contrast, offer a solution designed for small to medium experiments such as TAIGA. 
In such experiments, as a rule, historically developed specific data processing methods and specialized software are used.
We have specifically designed a distributed (cloud) data storage for astroparticle physics data collaboration in medium-sized experiments. In this article, we discuss the current state of our work using the example of the TAIGA and CASCADE experiments. 
A feature of our approach is that we provide our users with scientific data in the form to which they are accustomed to in everyday work on local resources.

\keywords{Astroparticle physics  \and Distributed data storage \and Metadata.}

\end{abstract}

\section{Introduction}

The modern physics of astroparticles is one of the most rapidly developing areas of modern science.
It includes several scientific fields, in each of which a number of large experimental installations have been put into operation.
So, in the field of gamma astronomy, the HESS [1] and MAGIC [2] experiments are working, and in 2023 the CTA installation [3] is to be commissioned.
In the field of neutrino astrophysics, we note IceCube~\cite{IceCube}, as well as the ongoing Global Neutrino Network (GNN)~\cite{GNN}, The Baikal deep underwater neutrino telescope (or Baikal-GVD - Gigaton Volume Detector)~\cite{Baikal}, which will include IceCube, KM3NeT~\cite{KM3NeT} and Baikal-GVD~\cite{Baikal}.
In the field of physics of high energies of cosmic rays, we note The Pierre Auger Observatory~\cite{Auger}.
The LIGO-Virgo consortium~\cite{LIGO,Virgo} sets a new direction in the study of gravitational waves.

However, in addition to the experimental megascience installations mentioned above, there are also medium and small installations.
An example of such installations is the TAIGA~\cite{TAIGA,TAIGA2}, TUNKA~\cite{TUNKA} installations deployed in the Tunkinskaya valley in Buryatia (Russia), the KASCADE~\cite{KASCADE2} installation and many others.

These installations also collect a large amount of data during their operation. An important feature of these experiments is that analysis of their data  is based on long-established practices using specific data processing methods and customized software. On the other hand, there is an insistent need to properly preprocess the collected data and make them accessible through a web service with a convenient and user-friendly interface. Putting it all together, it is important to provide web access to the data while maintaining the ability to work with the data using existing software and techniques.

This study was carried out within the framework of the Russian-German initiative~\cite{Bychkov18} aimed at supporting the processing of data from astrophysical experiments throughout the entire data life cycle: from collection and store to the preparation of data analysis results for publication and data archiving.

Of course, several approaches to the design of distributed data storages have already been proposed earlier. 
One of the most striking examples is the global system GRID~\cite{WLCG}, which was originally created to store and process LHC data, and later came to be used for many other experiments.

The International Virtual Observatory Alliance (IVOA)~\cite{IVOA} sets similar tasks. 
Another example is the CosmoHub~\cite{CosmoHub} system, which is based on the Hadoop distributed storage~\cite{Hadoop}.

All of these experiments use large project-oriented approaches that, for various reasons, may not be suitable for medium and small experiments.

Another important trend in astronomy is the combined analysis of data from various sources (multi-messenger astrophysics)~\cite{MMA19}, which is used to obtain a more detailed physical picture of the observed high energy astrophysical phenomena. 
In particular, a comparison of how the same phenomenon was observed by different small experiments could yield interesting new results. 
Making such a comparison requires the development of shared cloud storage for small experiments.

Thus, the development of cloud storage for small experts is an urgent task.

This article describes an approach to creating such a cloud storage and providing convenient access to data in it. 
The created cloud storage is called AstroDS.
The storage is focused on medium to small sized experiments such as TAIGA. 
The work is a logical continuation of the work of A.Kryukov with co-authors~\cite{Kryukov19}.

The structure of the article is as follows.
In Section 2, we provide a brief description of the principles that were used as the basis for the development of a data storage.
The third section is devoted to some peculiarities of working with remote storages based on data storage using relational databases using the example of KCDC~\cite{KCDC}.
The fourth section describes the AstroDS cloud storage prototype and its main characteristics.
In conclusion, we discuss the results obtained and the plan for the further development of the cloud storage.

\section{Design and architecture of the AstroDS}

As we said, AstroDS cloud storage is focused on small and medium experimental collaborations. This left a certain imprint on the decisions that were made during the system design process.

The main principle underlying the development of AstroDS cloud storage was the principle of maximizing preservation of the historically established methods of user interaction with local storages.
Thus, the main requirements were:
\begin{itemize}
\item preservation of the structure of data directories;
\item an opportunity to mount directories on local computers;
\item data transfer over the network should occurs at the time of real flicking to data.
\end{itemize}
This approach makes it possible to practically eliminate the modification of application software when working with cloud storage. 
At the same time, the load on the network is minimized.

Another very important goal was to make it as easy as possible to integrate existing local storages as a cloud storage node.
At the same time, the load on the local storage equipment should not significantly increase to ensure collaboration in the cloud storage.
To achieve this goal, an approach was chosen when all user requests are processed off-line on a special server that stores all the metadata necessary for data retrieval.
The collection of metadata is performed at the time of data loading to local storage.

Note that the storage implements a two-level data selection architecture:
\begin{itemize}
\item search at the file level, for example, by session number;
\item search at the level of individual events in files, for example, by the energy of the event.
\end{itemize}
This solution is flexible enough to fulfill almost any user request.

A special case is the integration of those local storage that store data on a per-life basis in a relational database.
An example of such a storage is the KCDC~\cite{KCDC} Data storage of the KASCADE~\cite{KASCADE2} Experiment. 
This case will be considered in more detail below.

In AstroDS all user requests are processed asynchronously, which can be important in some cases when you need to prepare a large sample.

Both the web interface and the command line interface are available to users.
The latter mode is more convenient when using a set of scripts to automate the data processing.

\begin{center}
\begin{figure}[h]
  \centering
  \begin{minipage}{1\textwidth}
    \includegraphics[width=\textwidth]{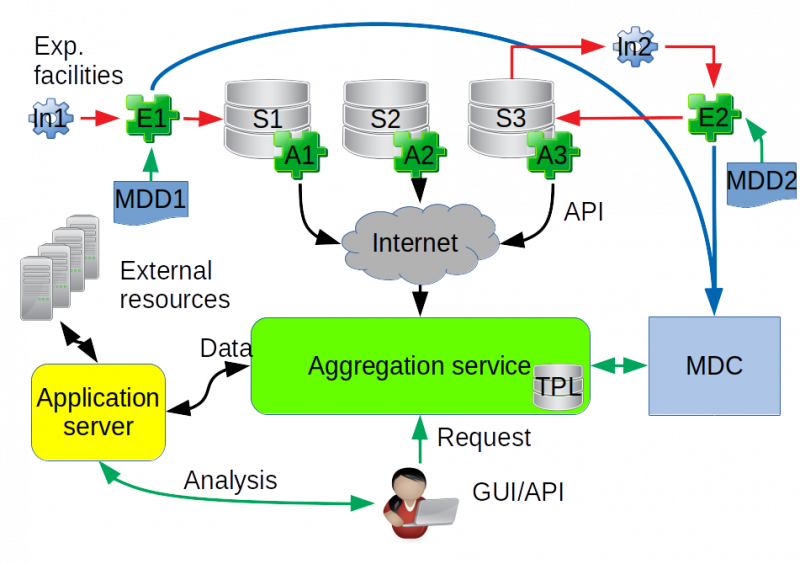}
    \caption{Simplified architecture of AstroDS.} 
    \label{fig1}
  \end{minipage}
\end{figure}
\end{center}

Taking into account all the above requirements, we have developed an architectural solution shown in Fig.~\ref{fig1}.

Below we will focus on a number of individual features of the implementation of the developed architecture. 
A more detailed presentation of the general principles of building the AstroDS system can be found in the articles ~\cite{Bychkov18,Kryukov19}.

\section{Metadata catalog}

The Metadata Catalog (MDC) is a single place where the physical location of the requested data is determined.
The MDC is a service which supports two main functions:
\begin{itemize}
    \item register collected metadata;
    \item process the user requests for data.
\end{itemize}

\subsection{Metadata catalog API}

\begin{center}
    \begin{figure}[h]
    \centering
        \begin{minipage}{1\textwidth}                                
        \includegraphics[width=1\linewidth]{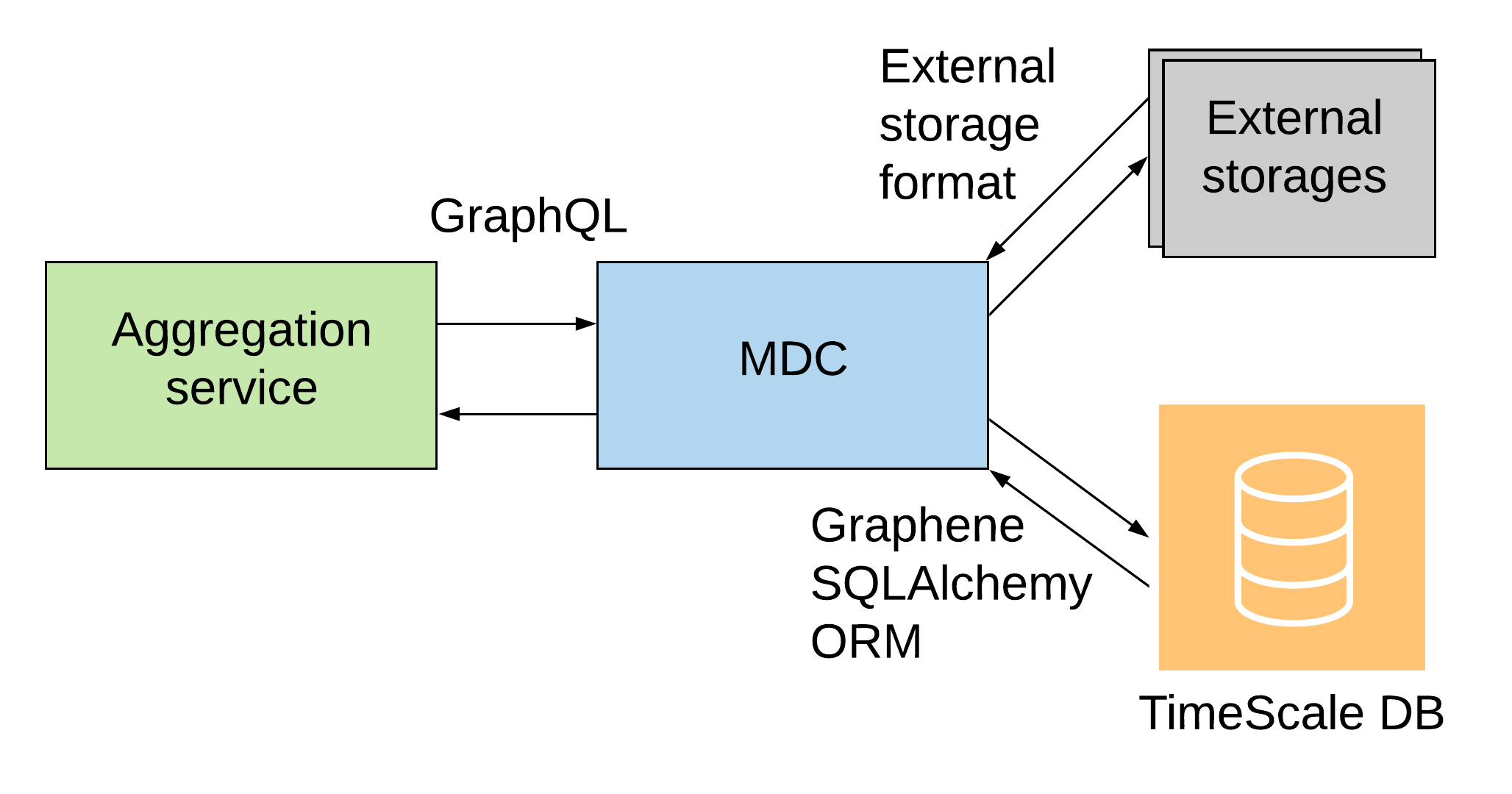}
        \caption{The architecture design of the MDC}
        \label{test1}
        \end{minipage}
    \end{figure}
\end{center}

The MDC architecture is based on the integration of several standard solutions (see Fig.~\ref{test1}). 
To store metadata we chose Timescale DB~\cite{timescale} -- a special database for storing time-series data. 
Flask~\cite{flask} is used as a web server for user requests processing. 
To enable the aggregation service to interact with the MDC, an API was implemented using GraphQL~\cite{graphql} query language. 
We used the Graphene-Python library~\cite{graphene_pyton} to easily create GraphQL APIs in Python.
SQL Alchemy as object relation mapper for TimeScale DB.

MDC provides an API for data insertion and for searching using the filter list shown in Table~\ref{mdc_par}. 
We do not provide an API for updating data in storage and deleting data from storage because the main idea of APPDS is that metadata is extracted only once by a special program called an extractor. 
All insert operations are implemented by a special GraphQL type - mutation.

\begin{table}[ht]
\centering
\begin{tabular}{|l|l|}
\hline
\textbf{Parameter name} & \textbf{Description} \\ \hline
fid                     & Unique file identifier                    \\ \hline
startTime               & Event start time                    \\ \hline
endTime                 & Event end time                    \\ \hline
first                   & Count of data for pagination                    \\ \hline
offset                  & Start position for pagination                    \\ \hline
weatherId               & Weather at the time of observation                    \\ \hline
trackingSourcId         & Observed object
                    \\ \hline
facilityId              & The facility that captured the event                    \\ \hline
\end{tabular}
\caption{Avaliable MDC parameters}
\label{mdc_par}
\end{table}

The query structure shown in Listing~\ref{query} consists of two main parts - data fields and query parameters. 
The data fields correspond to the DB schema and include such information as the run date, cluster, weather, facility, etc. 
All these data fields are primarily intended for the aggregation service, the end-user is interested in the url to download the file. 
The list of available query parameters allows you to filter data by event start and end time, facility, weather, tracking source, and unique file identifier. 

\begin{minipage}{1\textwidth}
\begin{lstlisting}[caption={The query structure} \label{query}]
query{
  files([query parameters]){
    [data fields]
  }
}
\end{lstlisting}
\end{minipage}

\begin{figure}[h]
    \includegraphics[width=1\linewidth]{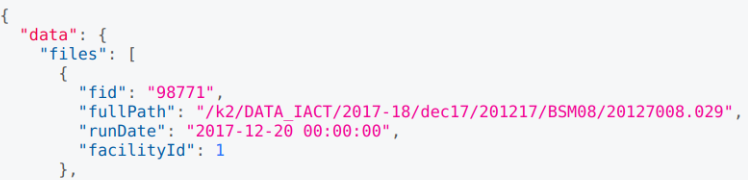}
    \caption{Example of GraphiQL response}
    \label{fig:graphiql}
\end{figure}

A GraphQL integrated development environment was deployed to test the API through the web GUI.
An example of the GraphQL response is shown in Fig.~\ref{fig:graphiql}.

\subsection{Filters for data selection}

The MDC can store data from different facilities. 
Each facility has its own parameters for filtering data.
In order for the aggregation service to form a list of filters for the client, the formal specifications for each facility are stored in the metadata catalog. 
Specifications include a list of options available for a given facility. 
Parameters can have one of five data types: "date", "int", "float", "string", "list". 
The first four parameters are the base data types. 
Type "list" could be a query string or an array of base types. 
The query string is required when the filtering value is contained in the database table. 
In this case, the aggregation service makes a request by this query string to the metadata catalog to get a list of parameters.

Specifications are stored in JSON format for each facility. In Listing~\ref{lst:filter} an example of the specification is shown. 
The start time and end time have "date" tipe and comparison sign as equal. 
The third parameter "weather" has type "list" and contains querying string. 

\begin{lstlisting}[caption={Filters specification }\label{lst:filter}] 
      {
        "filters": "[
           {\"name\": \"startTime\", \"type\": \"datetime\", 
            \"conditions\": [\"=\"]}, 
           {\"name\": \"endTime\",   \"type\": \"datetime\", 
            \"conditions\": [\"=\"]}, 
           {\"name\": \"weather\",   \"type\": \"list\",
            \"options\": { \"table\": \"weather\", 
               \"request\": \"query{weather{id wScale}}\", 
               \"fields\":  {\"id\":  {\"type\": \"integer\"}, 
                   \"wScale\":  {\"type\": \"string\"}}, 
               \"conditions\": [\"=\"]
            }
           }
         ]"
      }
\end{lstlisting}

\section{Integration with KCDC}

Sometimes access to raw data is not available. In this case, there is no way to extract the metadata and save it to the metadata catalog. 
For such cases, MDC is used as a proxy for user requests. 
In this case the third-party storages get request directly and process it themselves.

If third-party storages does not have a compatible query format with MDC API the MDC converts it in proper format and wise versa. 
So the aggregation service works with a uniform query format independently of storage API. 
For this purpose the only new converter should be added to MDC. 
This module describes how to translate formats to each other.

MDC knows all available external storages and their API formats. 
When MDC gets a request from aggregation service it translates GraphQL request to external request format and sends it and waits for the response. 
After, it translates the response back to GraphQL and sends it to aggregation service.

One example of how it works is shown in the Listing~\ref{lst:graphql} and Listing~\ref{lst:jsonrpc}. 
For this request, where facility id means request KASCAD data, fields start time and end time are converted from date-time format to timestamp. 
The request is formed in JSON with additional fields. 
And after receiving the response, MDC prepares a unified response for the aggregation service with the URL to download the file.
The availability of the file for download is checked on the side of the aggregation service.

\begin{lstlisting}[caption={GraphQL request}\label{lst:graphql}] 
query{
   files(
      facilityId: 5, 
      startTime: "2017-12-20T00:00:00"
      endTime: "2017-12-21T00:00:00"
  ){
      url
      facilityId
      __typename
  }
}
\end{lstlisting}

\begin{lstlisting}[caption={JSON--RPC 2.0 request}\label{lst:jsonrpc}] 
{
   "id": "cf919bb3-7028-4238-9438-2601cbfde3a6",
   "jsonrpc": "2.0",
   "method": "new_task",
   "parameters": {
      "type": ['kascade'],
      "datetime_min": 1256481198,
      "datetime_max": 1256481201
    }
}
\end{lstlisting}

\section{The Aggregation service}

The aggregation service (see Fig.~\ref{fig1}) is the central service of the AstroDS system and a single point of user entry into the system.
Its main tasks are as follows:
\begin{itemize}
\item building of user requests and their transfer to MDC;
\item requesting data from the local storages based on the MDC advice and providing the data requested by the user in the form of files and / or mount point of the virtual file system;
\item selection of events that meet the criteria of user requests from the received files and the formation of new subsets based on them.
\end{itemize}

The main requirement when developing an aggregation service is to minimize network traffic and load on remote data stores.
For this, files stored on remote storages are mounted on the aggregation service as a virtual file system CVMFS~\cite{CVMFS}. This file system initiates real data transfer only at the moment of actual data access.

Generating subsets of events that meet the criteria from a user request is a tedious task.
Using the aggregation service for this purpose minimized interference with the work of local storages and relieve them of the unpredictable load on their resources.

The aggregation service is described in more detail in the work of Nguyen et al.~\cite{Nguyen20}.

\section{AstroDS testbed}

\begin{center}
\begin{figure}[ht]
  \centering
  \begin{minipage}{1\textwidth}
    \includegraphics[width=\textwidth]{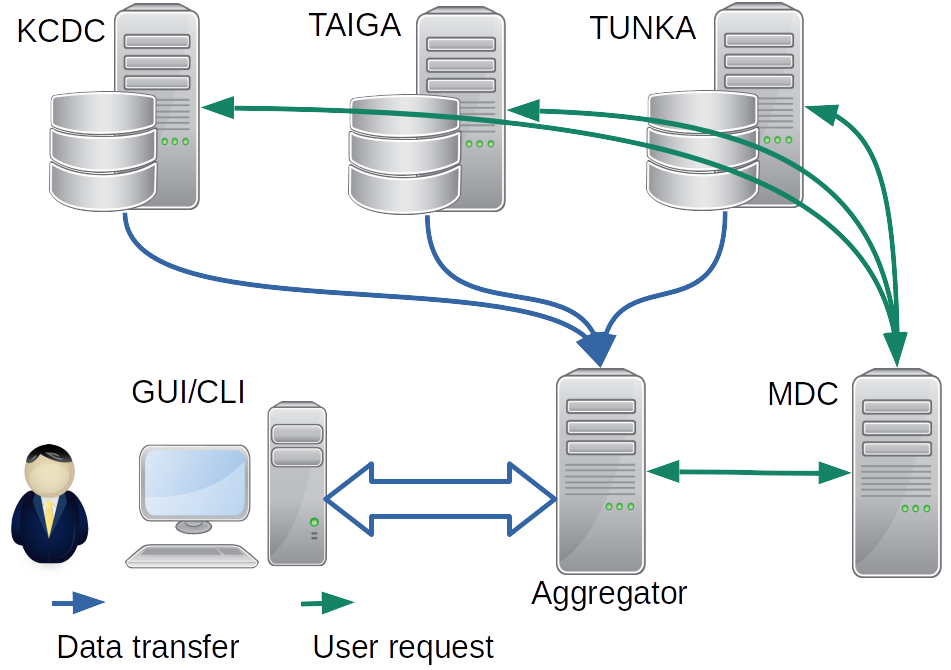}
    \caption{AstroDS test bed} 
    \label{tbed}
  \end{minipage}
\end{figure}
\end{center}

The AstroDS system is currently operating in a test mode at the test site, which includes three storage facilities. 
Two of them are real data storages of the TAIGA and KASCADE experiments, and one is the testers storage. 
Thus, the general structure of the polygon is as shown in Figure~\ref{tbed}.

The user can send requests through the aggregation service (Aggregator). 
The type of requests is still limited to file-level requests, but with the release of the update to version 2, the system will work both with file-level requests and requests that require a selection of individual events according to user criteria.
The main problem associated with event-level queries is the presence of specific meta-information, which, as a rule, arises after primary data processing or further analysis.

Note that among the three local storages integrated into the system, only the KCDC storage puts data event-by-event in a relational database.
This is reflected in the way we handle user requests.
Namely, since it is not known in advance whether there are events that meet the requirements in the user's request, then all requests, regardless of their content, are sent to KCDC and processed there.
As a response, KCDC always provides a link to a file (zip) that contains the selected events.
This file will be provided to the user either for download or for mounting on his work computer in the form of a virtual file system.
If there are no required events, KCDC will send a link to an empty file.

\section{Conclusion}

The article presents the results of the development of a distributed cloud data storage, called AstroDS, for medium and small astrophysical experiments.
It was shown that the principles underlying the construction of such a storage make it possible to practically exclude modification of application software and preserve the usual methods of working with data.

The deployed prototype of the AstroDS system currently includes data from the TAIGA, TUNKA and KASCADE experiments.

Integration of data from several experiments allows them to be used for joint analysis (multi-messenger), which will increase the accuracy of such analysis and the possibility of studying new phenomena.

In the future, it is planned to expand the functionality of the AstroDS cloud storage both in the direction of increasing the flexibility of data selection and in the number of experiments integrated into the system.

\section*{Acknolegement}

The authors are grateful to all RFN grant participants, as well as fellow participants of the Helmholtz Society Grant HRSF-0027. 
We would like to thank A. Haungs, V. Tokareva and J. Dubenskaya for fruitful discussions and assistance in preparing the manuscript.

%
%
%
%


\end{document}